%% file: main.tex
\documentclass[journal]{IEEEtran}
  
\IEEEoverridecommandlockouts                              

\usepackage{xr}

\newcommand{\omitted}[1]{}

\input{titleAndAuthors}

\usepackage{cite}

\usepackage{comment}
\usepackage{siunitx}
\usepackage{relsize}
\usepackage{ifthen}
\usepackage[colorinlistoftodos]{todonotes}

\usepackage[caption=false]{subfig}

\usepackage[vlined,ruled,linesnumbered]{algorithm2e}
\usepackage{graphics} 
\usepackage{rotating}
\usepackage{color}
\usepackage{enumerate}
\usepackage[T1]{fontenc}
\usepackage{psfrag}
\usepackage{epsfig} 
\usepackage{booktabs}
\usepackage{graphicx,url}
\usepackage{multirow}
\usepackage{array}
\usepackage{latexsym}
\usepackage{amsfonts}
\usepackage{amsmath}
\usepackage{amssymb}
\usepackage{xstring}
\usepackage{algorithmic}
\usepackage{multirow}
\usepackage{xcolor}
\usepackage{prettyref}
\usepackage{flexisym}
\usepackage{bigdelim}
\usepackage{breqn} 
\usepackage{listings}
\let\labelindent\relax
\usepackage{xspace}
\usepackage{bm}
\graphicspath{{./figures/}}
\usepackage{tikz}
\usetikzlibrary{matrix,calc}
\usepackage{lipsum}
\usepackage{mdwlist}

\let\itemize\undefined
\makecompactlist{itemize}{stditemize}
\usepackage{enumitem}
\usepackage{caption}
\usepackage{amsthm}
\usepackage{mathtools}

\makeatletter
\let\NAT@parse\undefined
\makeatother
\usepackage{hyperref}
\usepackage{cleveref}

\hypersetup{%
  colorlinks=true,
  linkcolor=blue,
  filecolor=magenta,      
  urlcolor=black,
  citecolor=red,
  linkbordercolor={0 0 1}
}

\input{preamble_symbols.tex}

\input{shortcuts.tex}

\PassOptionsToPackage{end}{algorithmic}

\renewcommand{\baselinestretch}{0.96}

\begin{document}








\maketitle

\input{0-Abstract.tex}
\input{1-Introduction.tex}
\input{2-RelatedWork.tex}
\input{3-Problem.tex}

\input{4-Algorithm.tex}

\input{5-Regret.tex}

\input{6-Exp.tex}
\input{7-Conclusion}

\input{Acknowledgements}

\appendices
\input{Appendix/Appendix.tex}

\bibliographystyle{IEEEtran}
\bibliography{References}



\end{document}

%% file: titleAndAuthors.tex
\title{
 \fontsize{19}{19} \selectfont 
Safe Non-Stochastic Control of Control-Affine Systems: \\ 
An Online Convex Optimization Approach
}

\author{
\thanks{Manuscript received June 24, 2023; Revised September 4, 2022; Accepted September 20, 2023. This paper was recommended for publication by Editor Jens Kober upon evaluation of the Associate Editor and Reviewers’ comments.}
Hongyu Zhou$^{1}$ \ Yichen Song$^{2}$ \ Vasileios Tzoumas$^{1}$
	\thanks{$^{1}$Department of Aerospace Engineering, University of Michigan, Ann Arbor, MI 48109 USA;  {\tt\footnotesize \{zhouhy, vtzoumas\}@umich.edu}}
	\vspace{-2mm}
    \thanks{$^{2}$Department of Mechanical Engineering, Boston University, Boston, MA 02215;  {\tt\footnotesize ycs@bu.edu}}
}

%% file: preamble_symbols.tex
\newtheorem{theorem}{Theorem}
\newtheorem{problem}{Problem}

\newtheorem{lemma}{Lemma}
\newtheorem{assumption}{Assumption}
\newtheorem{definition}{Definition}

\newtheorem{remark}{Remark}

\newcommand{\bdmath}{\begin{dmath}}
\newcommand{\edmath}{\end{dmath}}
\newcommand{\beq}{\begin{equation}}
\newcommand{\eeq}{\end{equation}}
\newcommand{\bdm}{\begin{displaymath}}
\newcommand{\edm}{\end{displaymath}}
\newcommand{\bea}{\begin{eqnarray}}
\newcommand{\eea}{\end{eqnarray}}
\newcommand{\beal}{\beq \begin{array}{lll}}
\newcommand{\eeal}{\end{array} \eeq}
\newcommand{\beas}{\begin{eqnarray*}}
\newcommand{\eeas}{\end{eqnarray*}}
\newcommand{\ba}{\begin{array}}
\newcommand{\ea}{\end{array}}
\newcommand{\bit}{\begin{itemize}}
\newcommand{\eit}{\end{itemize}}
\newcommand{\ben}{\begin{enumerate}}
\newcommand{\een}{\end{enumerate}}

\newcommand{\calC}{{\cal C}}
\newcommand{\calD}{{\cal D}}

\newcommand{\calK}{{\cal K}}

\newcommand{\calO}{{\cal O}}

\newcommand{\calS}{{\cal S}}

\newcommand{\calU}{{\cal U}}

\newcommand{\calW}{{\cal W}}

\definecolor{myblue}{RGB}{65 105 225}

\newcommand{\hide}[1]{}

\newcommand{\hiddenText}{{\color{gray} hidden text.}}
\newcommand{\hideWithText}[1]{\hiddenText}

\newcommand{\scenario}[1]{{\fontsize{9}{8.7}\selectfont\sf#1}\xspace}
\newcommand{\scenariot}[1]{{\fontsize{8}{8}\selectfont\sf#1}\xspace}



%% file: shortcuts.tex
\newcommand{\TimeSeq}{\{1, \dots, T\}}

\newcommand{\SumOneT}{\sum_{t=1}^{T}}

\newcommand{\ie}{\emph{i.e.},\xspace}
\newcommand{\eg}{\emph{e.g.},\xspace}

\newcommand{\red}[1]{{\color{red}#1}}
\newcommand{\blue}[1]{{\color{blue}#1}}

\newcommand{\myParagraph}[1]{{\bf #1.}\xspace}

\newcommand{\OGD}{\scenario{\textit{OGD}}}

\newcommand{\OCO}{\scenario{{OCO}}}

\newcommand{\SafeNSC}{\scenario{{Safe-NSC}}}
\newcommand{\SafeOGD}{\scenario{\textit{Safe-OGD}}}

\newcommand{\DDPG}{\scenario{\textit{DDPG}}}
\newcommand{\iLQR}{\scenario{\textit{iLQR}}}

\newcommand{\LF}{\scenario{\textit{LF}}}
\newcommand{\RNMPC}{\scenario{\textit{R-NMPC}}}

\newcommand{\DyRegNSC}{\operatorname{Regret-NSC}_T^D}

%% file: 0-Abstract.tex
\vspace{-5mm}
\begin{abstract}
We study how to safely control nonlinear control-affine systems that are corrupted with bounded \textit{non-stochastic noise}, \ie noise that is unknown a priori and that is \underline{not} necessarily governed by a stochastic model.  We focus on safety constraints that take the form of time-varying convex constraints such as collision-avoidance and control-effort constraints.  
We provide an algorithm with bounded \textit{dynamic regret}, \ie bounded suboptimality against an optimal clairvoyant controller that knows the realization of the noise a priori. 
We are motivated by the future of autonomy where robots will autonomously perform complex tasks despite real-world  unpredictable disturbances such as wind gusts.
To develop the algorithm, we capture our problem as a sequential game between a controller and an adversary, where the controller plays first, choosing the control input, whereas the adversary plays second, choosing the noise's realization.
The controller aims to minimize its cumulative tracking error despite being unable to know the noise's realization a priori.
We validate our algorithm in simulated scenarios of (i) an inverted pendulum aiming to stay upright, and (ii) a quadrotor aiming to fly to a goal location through an unknown cluttered environment.
\end{abstract}

\vspace{-2mm}

%% file: 1-Introduction.tex
\vspace{-4.5mm}
\section{Introduction}\label{sec:Intro}
\begin{figure}[t]
    \centering
    \includegraphics[width=0.4\textwidth]{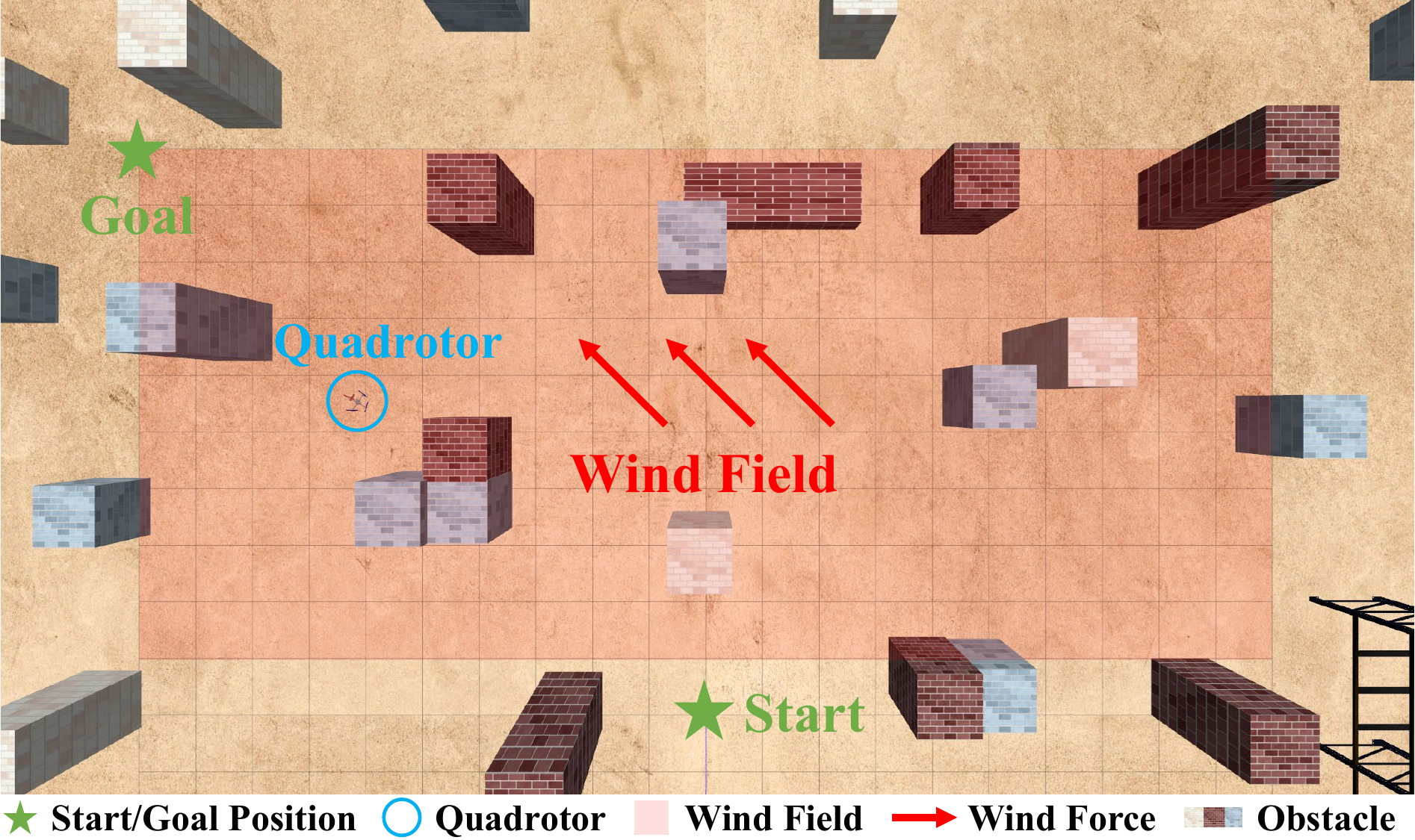}
    \captionsetup{font=footnotesize}
    \vspace{-1.5mm}
    \caption{\textbf{Safe non-stochastic control example: Autonomous flight in cluttered environments subject to unknown wind disturbances.} In this paper, we focus on safe non-stochastic control of control-affine systems where the robots' capacity to select effective control actions fast is challenged by (i) time-varying safety constraints, (ii) unknown, unstructured, and, more broadly, unpredictable noise, and (iii) nonlinear control-affine dynamics. For example, in package delivery with quadrotors, the quadrotors are required to fly to goal positions. But during such tasks,  (i) the quadrotors need to ensure collision avoidance at all times, which requires control actions that respect \textit{time-varying} state and control-input constraints, 
      (ii) the quadrotors may be disturbed by unpredictable wind gusts, 
     and (iii) they need to account for their nonlinear, in particular, control-affine dynamics.
     These challenges stress the quadrotors' ability to decide effective control inputs fast,  and to ensure safety.  
     We aim to provide a control algorithm that 
     handles these challenges,
     guaranteeing  bounded suboptimality against optimal safe controllers in hindsight. 
     }
	\vspace{-8mm}
    \label{fig_gazebo}
\end{figure}

\IEEEPARstart{I}{n} the future, robots will be leveraging their on-board control capabilities to complete safety-critical tasks such as package delivery~\cite{ackerman2013amazon}, target tracking~\cite{chen2016tracking}, and disaster response~\cite{rivera2016post}. To complete such complex tasks, the robots need to efficiently and {reliably} overcome a series of key challenges:

\vspace{.5mm}
\paragraph*{Challenge I: Time-Varying Safety Constraints} The robots need to ensure the safety of their own and of their surroundings.  For example, robots often need to ensure that they follow collision-free trajectories, or that their control effort is kept under prescribed levels. Such safety requirements take the form of {time-varying} state and control input constraints: \eg as robots move in cluttered environments, the current obstacle-free environment changes (\Cref{fig_gazebo}).  Accounting for such constraints in real-time can be challenging, requiring increased computational effort~\cite{rawlings2017model,borrelli2017predictive}.  Hence, several real-time state-of-the-art methods do not guarantee safety at all times~\cite{chen2017constrained,cosner2023robust}. 

\paragraph*{Challenge II: Unpredictable Noise} The robots' dynamics are often corrupted by unknown non-stochastic noise, \ie noise that is not necessarily i.i.d.~Gaussian or, more broadly, that is not governed by a \textit{stochastic} (probability) model.  For example, aerial and marine vehicles often face non-stochastic winds and waves, respectively~\cite{faltinsen1993sea}.  But the current control algorithms primarily rely on the assumption of known stochastic noise, typically, Gaussian, compromising the robots' ability to ensure safety in real-world settings where this assumption is violated~\cite{aastrom2012introduction}.

\paragraph*{Challenge III: Nonlinear, Control-Affine Dynamics} The dynamics of real-world robots are often nonlinear, in particular, control-affine. For example, the dynamics of quadrotors and marine vessels take the form of 
$x_{t+1} = f\left(x_{t}\right) + g\left(x_{t}\right) u_{t} +w_{t}$, where (i) $x_t$ is the robot's state, (ii) $f\left(x_{t}\right)$ and $g\left(x_{t}\right)$ are system matrices characterizing the robot's dynamics,  (iii) $u_{t}$ is the control input, and (iv) $w_{t}$ is the disturbance~\cite{khalil2002modeling}.
Handling nonlinear dynamics often requires complex control policies, requiring, in the most simple case, linearization at each time step, and, thus, additional computational effort~\cite{rawlings2017model}.

The above challenges motivate control methods that efficiently and reliably handle control-affine systems, and guarantee the satisfaction of time-varying safety constraints even in the presence of non-stochastic noise.
 State-of-the-art methods to this end
typically rely on robust control~\cite{mayne2005robust,goel2020regret,sabag2021regret,martin2022safe,didier2022system,zhou2022safe} or on online learning~\cite{agarwal2019online,simchowitz2020improper,li2021online,zhao2021non,boffi2021regret,zhou2023efficient,zhou2023safe}. But the robust methods are often conservative and computationally heavy: not only  they simulate the system dynamics over a lookahead horizon; they also assume a worst-case noise realization, given a known upper bound on the magnitude of the noise.
To reduce conservatism and increase efficiency, researchers have also focused on online learning methods via Online Convex Optimization (\OCO) \cite{hazan2016introduction}. These methods rely on the \textit{Online Gradient Descent} (\OGD) algorithm or its variants, offering bounded \textit{regret} guarantees, \ie bounded suboptimality with respect to an optimal (possibly time-varying) clairvoyant controller that knows the future noise realization a priori~\cite{agarwal2019online,simchowitz2020improper,li2021online,zhao2021non,zhou2023efficient}. However, the current online methods address only linear dynamical systems and time-\underline{in}variant safety constraints.


\myParagraph{Contributions} We provide an algorithm for the  problem of 
controlling control-affine systems, guaranteeing their safety against time-varying safety constraints even in the presence of non-stochastic noise.  Henceforth, we refer to this problem as \textit{Safe Non-Stochastic Control for Control-Affine Systems} (\SafeNSC).  Our solution approach, its generality, and its performance guarantees are as follows:

\paragraph{Approach} 
     We first formalize the problem of 
     \SafeNSC
     as a sequential game between a controller and an adversary, inspired by the current literature on online learning and control~\cite{agarwal2019online,simchowitz2020improper,li2021online,zhao2021non,zhou2023efficient,zhou2023safe}:
at each time step $t$, the controller plays first, choosing a control input, and,
then, the adversary plays second, choosing the noise's realization.
The controller aims to minimize its cumulative performance, \eg tracking error, despite being unaware of the noise’s realization a priori. 

We then provide an algorithm for \SafeNSC, called \textit{Safe Online Gradient Descent} (\SafeOGD) (\Cref{alg:SafeOGD_Control}).
\SafeOGD chooses control input $u_t$ online from a set $\calU_t$ that is also chosen by \SafeOGD online to guarantee safety.  In more detail, $\calU_t$ is constructed (i) given the current safety constraints, and an upper bound to the non-stochastic noise, and (ii) employing \textit{Control Barrier Functions}~(CBF)~\cite{ames2014control} to enforce the constraints.  The safety constraints may be known a priori or constructed online given the currently known environment, \eg the current obstacle-free environment within field of view.

\paragraph{Generality} \SafeOGD, as presented above, selects control inputs directly, without optimizing an underlying policy parameterization. Oftentimes, however, it is desired to optimize a control-policy parametrization, such as in Linear-Quadratic Gaussian (\scenario{LQG}) control~\cite{xue2007linear}. 
We show that \SafeOGD can be extended to optimize online any control-policy parameterization that is linear in the  parameters (\Cref{subsec:extention}). Examples of such policies include the linear state-feedback control policy~\cite{xue2007linear}, which is commonly used for \scenario{LQG} control, and the disturbance-action policy~\cite{borrelli2017predictive,agarwal2019online}, \mbox{which is commonly used for non-stochastic control.}

    \paragraph{Performance Guarantees and Near-Optimality in Classical Control Settings} We prove that the \SafeOGD controller has bounded dynamic regret against any clairvoyant control policy (\Cref{theorem:SafeOGD_Control}).
    The regret bound also implies (near-)optimal performance in classical online convex optimization and optimal control problems:  
    (i) when the domain of optimization is time-\underline{in}variant,
    \SafeOGD's performance bound reduces to that of the seminal \OGD algorithm for online convex optimization, which bound is near-optimal~\cite{zinkevich2003online};
    and (ii) when also the optimal clairvoyant control policy is time--\underline{in}variant, \SafeOGD learns asymptotically the optimal control policy (\Cref{sec:SafeOCO-Reg}), implying, for example, that \SafeOGD would converge to the optimal linear state-feedback controller if applied to the classical \scenario{LQG} problem.

\myParagraph{Numerical Evaluations} {
We validate our algorithms with extensive simulations (\Cref{sec:exp}). Specifically, we validate our algorithm in simulated scenarios of (i) an inverted pendulum that is tasked to stay upright, and (ii) a quadrotor that is tasked to move to a prescribed location  by flying through an unknown cluttered environment (Fig.~\ref{fig_gazebo}).  We conduct these experiments in Python and Gazebo, respectively. In the inverted pendulum experiment (\Cref{subsec:sim-1}), we compare our algorithm with a linear feedback controller, the \textit{Deep Deterministic Policy Gradient} (\DDPG) controller~\cite{silver2014deterministic,lillicrap2015continuous}, and the \textit{iterative Linear–Quadratic Regulator} (\iLQR)~\cite{chen2017constrained}. 
In the quadrotor experiment (\Cref{subsec:sim-2}), we compare our algorithm with the geometric controller~\cite{lee2010geometric} and a \textit{Robust Nonlinear Model Predictive Controller} (\RNMPC)~\cite{wu2021external}.

Our algorithm achieves in the simulations a better performance, achieving (i) safety at all times, whereas the \iLQR and geometric controllers as well as \RNMPC do not, and (ii) better or comparable tracking performance than the linear feedback, \DDPG, and geometric controllers as well as \RNMPC.

%% file: 2-RelatedWork.tex
\vspace{-4mm}
\section{Related Work}\label{sec:lit_review}

The said Challenges I to III have motivated research on robust control, online learning for control, and safe control:

\myParagraph{Robust control}
Robust control algorithms select control inputs upon simulating the future system dynamics across a {lookahead} horizon~\cite{hassibi1999indefinite,goel2020regret,sabag2021regret,martin2022safe,didier2022system,zhou2022safe}. To this end, they either assume a worst-case realization of noise, given an upper bound to the magnitude of the noise~\cite{goel2020regret,sabag2021regret,martin2022safe,didier2022system,zhou2022safe}, or assume a stochastic model governing the evolution of the noise~\cite{hassibi1999indefinite}.
However, assuming the worst-case noise realization can oftentimes be pessimistic; and assuming a stochastic model may compromise performance when the underlying noise is non-stochastic or the assumed model is incorrect.

\myParagraph{Online learning for control} Online learning algorithms select control inputs based on past information only since they assume no model that can be used to simulate the future evolution of the noise~\cite{agarwal2019online,li2021online,simchowitz2020improper,zhao2021non,boffi2021regret,zhou2023efficient,zhou2023safe}. Assuming a known upper bound to the magnitude of the noise, and by employing the \OCO framework to capture the non-stochastic control problem as a sequential game between a controller and an adversary~\cite{hazan2016introduction}, they provide bounded regret guarantees against an optimal (time-varying) clairvoyant controller even under unpredictable noise.  They consider time-\underline{in}variant state and control input constraints {or no constraint}, in contrast to time-varying safety constraints, with the exception of~\cite{zhou2023safe}. Also, they only consider linear dynamical systems, in contrast to nonlinear control-affine systems, {with the exception of~\cite{boffi2021regret}}.\footnote{We compare this paper with our previous work~\cite{zhou2023safe}: (i) \cite{zhou2023safe} considers linear time-varying systems; in contrast, herein we consider nonlinear control-affine systems. (ii) \cite{zhou2023safe} employs toy MATLAB simulations of a linear time-invariant model of a quadrotor that aims to stay at the hovering position; in contrast, we conduct extensive experiments on two nonlinear control-affine systems, namely, an inverted pendulum aiming to stay upright, and a quadrotor flying in an unknown cluttered environment, where we use Python and Gazebo, respectively. (iii) \cite{zhou2023safe} is accepted to a conference venue and will not contain any proof of its theoretical results due to space limitations. (iv)  There is no overlap in the writing between \cite{zhou2023safe} and this paper.}

\myParagraph{Safe Control}
Safe control algorithms select control inputs to ensure that the unsafe region on the system’s state space is not reachable. This can be achieved by using Hamilton-Jacobi (HJ) reachability analysis \cite{bansal2017hamilton} and CBF~\cite{ames2014control}. HJ reachability analysis computes the backward reachable set, \ie the set of states from which the system can be driven into the unsafe set, and selects control inputs to avoid
such an unsafe set. HJ reachability analysis is able to handle bounded disturbances and recovers maximal safe sets. However, it is often computationally intractable for high-dimensional systems, thus, requiring to place assumptions on system dynamics to recover traceability, 
such as requiring linear systems or the system dynamics can be decomposed into several subsystems~\cite{maidens2013lagrangian,darbon2016algorithms,herbert2021scalable}. 
CBF achieves safety by ensuring forward invariance of the safe set, \ie by selecting control inputs such that the system state remains in the safe set. {This method has been used to design control algorithms for systems with no noise~\cite{ames2014control,ames2016control,agrawal2017discrete,zeng2021safety}, stochastic noise~\cite{clark2019control,cosner2023robust}, and bounded non-stochastic noise~\cite{jankovic2018robust}.}

%% file: 3-Problem.tex
\vspace{-3mm}
\section{Problem Formulation}\label{sec:problem}

We formulate the problem of \textit{Safe Non-Stochastic Control for Control-Affine Systems} (\Cref{prob:SafeNSC}).  To this end, we use the following framework and assumptions.

\myParagraph{Control-Affine Systems}
We consider discrete-time control-affine systems of the form
\vspace{-2.5mm}
\begin{equation}
	x_{t+1} = f\left(x_{t}\right) + g\left(x_{t}\right) u_{t} +w_{t}, \quad t \in \{1, \ldots, T\}, 
    \label{eq:affine_sys}
\vspace{-1.5mm}
\end{equation}
where $x_t \in\mathbb{R}^{d_x}$ is the state, $u_t \in\mathbb{R}^{d_u}$ is the control input, $w_t\in\mathbb{R}^{d_x}$ is the process noise, and $f: \mathbb{R}^{d_x} \blue{\rightarrow} \mathbb{R}^{d_x}$ and $g: \mathbb{R}^{d_x} \blue{\rightarrow} \mathbb{R}^{d_x} \times \mathbb{R}^{d_u}$ are known locally Lipschitz functions. 

For simplicity, we assume the stability of the nominal system in \cref{eq:affine_sys} without the noise:
\vspace{-1.5mm}
\begin{assumption}[Stability of the Nominal System]\label{assumption:stability}
The nominal system, $x_{t+1} = f\left(x_{t}\right) + g\left(x_{t}\right) u_{t}$, is stable.
\end{assumption}
\vspace{-3mm}
\begin{remark}[Removal of the Stability Condition]
The stability condition can be removed by employing a known stabilizing controller $u_{t}^{s}$, that is, setting $u_t = u_{t}^{s} + v_{t}$, where the stabilizing controller $u_{t}^{s}$ can be computed via Control Lyapunov Functions~\cite{khalil2015nonlinear}, and $v_{t}$ is the controller to be designed.
\end{remark}
\vspace{-1.5mm}
For a quadrotor, such a stabilizing controller can be one that enables the quadrotor to track a given reference path~\cite{sun2022comparative}.
\vspace{-1.5mm}
\begin{assumption}[Bounded Noise]\label{assumption:bounded_system_noise}
The noise is bounded, \ie $w_t \in \calW \triangleq \{w \mid \left\|w\right\| \leq W\}$, where $W$ is given.
\end{assumption}
\vspace{-1.5mm}
Per \Cref{assumption:bounded_system_noise}, we assume no stochastic model for the process noise $w_t$: the noise may even be adversarial, subject to the bound $W$. 
An example is a quadrotor subject to wind disturbances with bounded magnitude, whose evolution may not be governed by a known stochastic model.

\myParagraph{Safety Constraints} We consider the states and control inputs for all $t$ must satisfy polytopic constraints of the form\footnote{Our results hold true also for any constraints such that \cref{eq:DCBF} is convex  $u_t$. We focus on polytopic constraints for simplicity in the presentation.}
\vspace{-2.5mm}
\begin{equation}
    \begin{aligned}
        & x_t \in \calS_t \triangleq \{x \mid L_{x,t} x \leq l_{x,t}\}, \ \forall \{{w}_{\tau} \in \calW\}_{\tau=1}^{t-1}, \\
        & u_t \in \calC_t \triangleq \{u \mid L_{u,t} u \leq l_{u,t}\}, 
    \end{aligned}
    \label{eq:safety_constraints}
\vspace{-1.5mm}
\end{equation}
where $L_{x,t}$, $l_{x,t}$, $L_{u,t}$, and $l_{u,t}$ are given.
\vspace{-1.5mm}
\begin{assumption}[Bounded State and Control Input Domain]\label{assumption:bounded_set}
    The  domain sets $\calS_t$ and $\calC_t$ are bounded for all $t\in\TimeSeq$. Also, they are contained in the bounded sets $\mathcal{S}$ and $\calC$, respectively, that both contain the zero point and have diameter $D$.\footnote{A set $\calS$ contains the zero point and has \textit{diameter} $D$ when $\mathbf{0} \in \calS$, and $\|x-y\|\; \leq D$ for all $x \in \mathcal{S}, \mathbf{y} \in \mathcal{S}$.}
\end{assumption}
\vspace{-1.5mm}
\Cref{assumption:bounded_set} ensures that the loss function has a bounded gradient with respect to the state and control input; this helps to quantify \SafeOGD's dynamic regret. Bounded state constraints emerge, for example, when a quadrotor flies in a cluttered environment (Fig.~\ref{fig_gazebo}): its state must be in a bounded flight corridor to ensure safety. Bounded control input constraints are needed to avoid controller saturation.

\vspace{-1.5mm}
\begin{assumption}[Existance of Safe Controller]\label{assumption:safe}
    Given $x_t \in \calS_t$, there exists a safe controller $u_t \in \calC_t$ such that $x_{t+1}\in \calS_{t+1}$, $\forall w_t \in \calW$.
\end{assumption}
\vspace{-1.5mm}
\Cref{assumption:safe} enables the constraint $L_{x,t} x \leq l_{x,t}$ to be used as a CBF, introduced in \Cref{subsec:DCBF}. 
This assumption is reasonable if the robot is kept at a low speed and the control authority is large enough such that the robot is able to react quickly and keep safe even if it is close to an obstacle. 
In general, the constraint $L_{x,t} x \leq l_{x,t}$ must be constructed by taking into account the robot's speed and acceleration capacity such that there always exists a control input $u_t$ that satisfies $x_{t+1} \in \calS_{t+1}$, an example of which is given in \cite{ames2014control}.

\myParagraph{Control Performance Metric} We design the control inputs $u_t$ to ensure: (i) safety; and (ii) a control performance that is comparable to an optimal clairvoyant policy that knows the future noise realizations ${w}_1, w_2, \ldots$ a priori. Particularly, we consider the control performance metric defined below.

\vspace{-1.5mm}
\begin{definition}[Dynamic Regret]\label{def:DyReg_control}
Assume a lookahead time horizon of operation $T$, and loss functions $c_t$, $t=1,\ldots, T$. Then, \emph{dynamic regret} is defined as
\vspace{-2.5mm}
\begin{equation}
	\DyRegNSC = \sum_{t=1}^{T} c_{t}\left(x_{t+1}, u_{t}\right)-\sum_{t=1}^{T} c_{t}\left(x_{t+1}^{*}, u_{t}^{*}\right),
	\label{eq:DyReg_control}
        \vspace{-2mm}
\end{equation}
where (i) both sums in \cref{eq:DyReg_control} are evaluated with the same noise $\{w_1,\ldots, w_T\}$ that is the noise experienced by the system during its evolution per the control sequence $\{u_1,\ldots, u_T\}$, (ii) $u_{t}^{*}$ is the optimal control input in hindsight, \ie the optimal input given a priori knowledge of the noise $w_t$,  which includes the control inputs generated by a nonlinear control policy, (iii) $x_{t+1}^{*}$ is the state reached by applying the optimal control inputs $u_{t}^{*}$ from state $x_{t}$, which is the optimal state could have been reached when the system is at $x_{t}$, and (iv) $x_{t+1}^{*}$ and $u_{t}^{*}$ satisfy the constraints in \cref{eq:safety_constraints} for all $t$.
\end{definition}

\vspace{-3.5mm}
\begin{assumption}[Convex and Bounded Loss Function, and with Bounded Gradient]\label{assumption:cost}
$c_{t}\left(x_{t+1},u_t\right): \mathbb{R}^{d_{x}} \times \mathbb{R}^{d_{u}} \blue{\rightarrow} \mathbb{R}$ is convex in $x_{t+1}$ and $u_t$. 
Further, when $\|x\|$ and $\|u\|$ are bounded, then $\left|c_{t}(x, u)\right|$, $\left\|\nabla_{x} c_{t}(x, u)\right\|$, and $\left\|\nabla_{u} c_{t}(x, u)\right\|$ are also bounded.
\end{assumption} 
\vspace{-1.5mm}
A loss function that satisfies \Cref{assumption:cost} is the commonly used quadratic cost $c_{t}\left(x_{t+1}, u_{t}\right) = x_{t+1} Q x_{t+1}^\top + u_{t} R u_{t}^\top$.

\myParagraph{Problem Definition} We formally define the problem of \textit{Safe Non-Stochastic Control of Control-Affine Systems}:
\vspace{-1.5mm}
\begin{problem}[Safe Non-Stochastic Control of Control-Affine Systems (\SafeNSC)]\label{prob:SafeNSC}
Assume that the initial state of the system is safe, that is, $x_1 \in \calS_{1}$. At each $t=1,\ldots, T$, identify a control input $u_t$ that guarantees safety, that is, $x_t\in\calS_t$ and $u_t\in\calC_t$, and such that $\DyRegNSC$ is minimized by time $T$.
\end{problem}

%% file: 4-Algorithm.tex
\vspace{-4mm}

\section{Algorithm for Safe Non-Stochatic Control}\label{sec:alg}

We present the \textit{Safe Online Gradient Descent} (\SafeOGD) algorithm (\Cref{alg:SafeOGD_Control}).
We first provide background information on control barrier functions since \SafeOGD utilizes them to ensure safety (\Cref{subsec:DCBF}). Then, we present the basic version of \SafeOGD that directly optimizes the control inputs (\Cref{subsec:SafeOGD}).  Finally, we discuss how the algorithm is extended to optimize linear control policies (\Cref{subsec:extention}).

\vspace{-4mm}
\subsection{Preliminaries: Discrete-Time Control Barrier Functions}\label{subsec:DCBF}
Consider a smooth function $h: \mathbb{R}^{d_x} \blue{\rightarrow} \mathbb{R}$, a safety set $\mathcal{D}$ and its boundary defined as $\mathcal{D}  \triangleq \{x \in \mathbb{R}^{d_x} \mid h\left(x\right) > 0\}$, $\partial \mathcal{D}  \triangleq \{x \in \mathbb{R}^{d_x} \mid h\left(x\right)=0\}$.

\vspace{-2mm}
\begin{definition}[Discrete-Time Exponentially Control Barrier Function) (DCBF) \cite{agrawal2017discrete}]
A map $h: \mathbb{R}^{d_x} \blue{\rightarrow} \mathbb{R}$ is a \emph{Discrete-Time Exponentially Control Barrier Function} on $\calD$ if 
there exists a control input $u_{t}\in\calC_t$ and a positive constant $\alpha~\in~(0, 1]$ such that $\Delta h_{t} \triangleq h\left(x_{t+1}\right) - h\left(x_{t}\right)$ satisfies
    \vspace{-1.5mm}
            \begin{equation}
                \Delta h_{t} + \alpha h\left(x_{t}\right) \geq 0.
                \label{eq:DCBF}
                \vspace{-1.5mm}
            \end{equation}
\end{definition}

The value of DCBFs in control is that they can be used to ensure safety: a control input $u_{t}$ renders the
safety set $\mathcal{D}$ forward invariant when \cref{eq:DCBF} is satisfied. 


\vspace{-5mm}
\subsection{\SafeOGD Algorithm: The Basic Algorithm}\label{subsec:SafeOGD}

\input{Alg/Alg-SafeOGD-Control.tex}

We present \SafeOGD in \Cref{alg:SafeOGD_Control}. The algorithm first initializes $u_1 \in \calU_1$ (line 1). At each iteration $t$, \Cref{alg:SafeOGD_Control} evolves to state $x_{t+1}$ applying the control inputs $u_t$ (line 3) and calculates the noise $w_t$ upon observation of $x_t$ (line 4). After that, the algorithm suffers a loss of $c_t\left(x_{t+1},u_t\right)$ (line 5). Then, \Cref{alg:SafeOGD_Control} expresses $c_t\left(x_{t+1},u_t\right)$ as a function of $u_t$, \ie $c_t(u_t) \triangleq c_t(x_{t+1}, u_t) = c_t(f(x_{t}) + g(x_{t}) u_{t} + w_{t}, u_t)$ ---which is convex in $u_t$, given functions $f\left(\cdot\right)$ and $g\left(\cdot\right)$, $x_t$, and $w_t$, per \Cref{lemma:cvx_u} below--- and obtains the gradient $\nabla_t \triangleq \nabla_u c_t\left(x_{t+1}, u_t\right)$ (lines 6-7). To ensure safety, \Cref{alg:SafeOGD_Control} constructs the domain set $\calU_{t+1}$ per \Cref{lemma:u_t} (line 8). Finally, \Cref{alg:SafeOGD_Control} updates the control gain and projects it back to the domain set $\calU_{t+1}$ (lines 9-10).

\vspace{-2mm}
\begin{lemma}[Convexity of Loss function in Control Input]\label{lemma:cvx_u}
The loss function $c_{t}\left(x_{t+1}, u_{t}\right) : \mathbb{R}^{d_x} \times \mathbb{R}^{d_u} \blue{\rightarrow} \mathbb{R}$ is convex in the control input $u_t$, given functions $f\left(\cdot\right)$ and $g\left(\cdot\right)$, $x_t$, and $w_t$.
\end{lemma}
\vspace{-3mm}
\begin{lemma}[Construction of Time-Varying Domain Set with Safety Guarantee]\label{lemma:u_t}
\Cref{alg:SafeOGD_Control} guarantees $x_{t+1} \in \calS_{t+1}$ and $u_t \in \calC_t$ at each time step $t$ by choosing $u_t \in \calU_t$, where
\vspace{-2mm}
\begin{equation}
\hspace*{-2mm}
    \begin{aligned}
    \calU_t \triangleq  \{ u \mid & \   L_{u,t} u \leq l_{u,t}, \\
                    & - L_{x,t+1} g\left(x_{t}\right) u - \left\| L_{x,t+1} \right\| W + l_{x,t+1} \\
                    &   - L_{x,t+1} f\left(x_{t}\right) - (1-\alpha)\left(l_{x,t} - L_{x,t} x_t\right) \geq 0 \}.
    \end{aligned}
    \label{eq:lemma_u_t}
\end{equation}
\vspace{-2mm}
\end{lemma}
\vspace{-2mm}
The origin of the two inequalities in \cref{eq:lemma_u_t} are as follows: the first condition is due to the requirement $u_t \in \calC_t$; and the second condition is the result of applying the DCBF condition in \cref{eq:DCBF} to guarantee $x_{t+1} \in \calS_{t+1}$.


\vspace{-4mm}
\subsection{Extension of \SafeOGD to Linear Control Policies}\label{subsec:extention}

We extend \Cref{alg:SafeOGD_Control} to optimize any linearly parameterized control policy in the form of control parameters multiplying state or noise, \eg the linear state-feedback policy~\cite{xue2007linear} and the disturbance-action policy~\cite{borrelli2017predictive,agarwal2019online}:

\begin{itemize}[leftmargin=9pt]
    \item \textit{Linear State-Feedback Control Policy:}
    This policy takes the form of $u_t = - K_t x_t$, where $K_t \in \mathbb{R}^{d_u \times d_x}$ is the control parameterization, and $\|K_t\|\leq \kappa$ with $\kappa > 0$.
    \item \textit{Disturbance-Action Control Policy:}
    This policy is defined as $u_t = \sum_{i=1}^{H} K_t^{[i]} w_{t-i}$, where  $K_t = (K_t^{[1]}, \dots, K_t^{[H]}) \in \mathbb{R}^{H \times d_u \times d_x}$ is the control parameterization, $H$ is a positive integer, and $\|K_t^{[i]}\| \leq \kappa$.
\end{itemize}

The extension of \Cref{lemma:cvx_u} follows trivially as $u_t$ is linear in $K_t$. The extension of \Cref{lemma:u_t} follows by substituting $u_t$ with the choice of control policy, \eg $ - K x_t$, and  $\sum_{i=1}^{H} K^{[i]} w_{t-i}$.

\myParagraph{Modification of \Cref{alg:SafeOGD_Control} to Learning Linearly Parameterized Control Policies}
To enable \Cref{alg:SafeOGD_Control} to learn linear control policies, the following modifications are needed: 
\begin{itemize}[leftmargin=9pt]
    \item The domain sets $\calU_t$ (lines 1, 8, 10) should be changed to $\calK_t$, where $\calK_t$ is the set of control parameterization that ensures safety, calculated based on \Cref{lemma:u_t} for $K_{t}$;
    \item The loss function $c_t$ (line 6) and the gradient (line 7) should be respected to the control parameterization $K_t$;
    \item The update and project steps (lines 9-10) are performed on the control parameterization $K_t$.
\end{itemize}

%% file: Alg/Alg-SafeOGD-Control.tex
\setlength{\textfloatsep}{-0.1mm}
\begin{algorithm}[t]
\small
	\caption{Safe Non-Stochastic Control Algorithm for Control-Affine Systems.}
	\begin{algorithmic}[1]
		\REQUIRE Time horizon $T$; gradient descent step size $\eta$.
		\ENSURE Control ${u}_t$ at each time step $t=1,\ldots,T$.
		\medskip
        \vspace{-1.5mm}
        \STATE Initialize ${u}_1 \in \calU_1$; 
		\FOR {each time step $t = 1, \dots, T$}
		\STATE Apply control input $u_t$;
        \STATE Observe the state $x_{t+1}$, and calculate the noise $w_t = x_{t+1} - f(x_{t}) - g(x_{t}) u_{t}$;
        \STATE Suffer the loss $c_t(x_{t+1}, u_t)$;
        \STATE Express the loss function in $u_t$ as $c_t(u_t) \triangleq c_t(x_{t+1}, u_t) = c_t(f(x_{t}) + g(x_{t}) u_{t} + w_{t}, u_t)$;
		\STATE Calculate gradient $\nabla_t \triangleq \nabla_u c_t(x_{t+1}, u_t)$;
		\STATE Calculate domain set $\calU_{t+1}$ per \cref{eq:lemma_u_t};
		\STATE Update $u_{t+1}^\prime=u_t- \eta \nabla_t$;
		\STATE Project  $u_{t+1}^\prime$ onto $\calU_{t+1}$, \ie $u_{t+1} = \Pi_{\calU_{t+1}}(u_{t+1}^\prime)$;
		\ENDFOR
	\end{algorithmic}\label{alg:SafeOGD_Control}
\end{algorithm}

%% file: 5-Regret.tex
\vspace{-4mm}
\section{Dynamic Regret Analysis}\label{sec:SafeOCO-Reg}

We present the dynamic regret bounds of \SafeOGD, both for the basic algorithm and for its extension to learning linearly parameterized control policies. 
We use the notation:
\begin{itemize}[leftmargin=9pt]
    \item $\bar{u}_{t+1} \triangleq \Pi_{\calU_{t}}(u_{t+1}^\prime)$ is the control would have been chosen at time step $t+1$ if $\calU_t = \calU_{t+1}$;

    \item $\zeta_t \triangleq \left\|\bar{u}_{t+1} - {u}_{t+1} \right\|$ 
    is the distance between $\bar{u}_{t+1}$ and ${u}_{t+1}$, which are the projection of ${u}_{t+1}^\prime$ onto sets $\calU_t$ and $\calU_{t+1}$, respectively.  Thus, it quantifies how fast the safe domain set changes ---$\zeta_t$ is  $0$ when $\calU_t=\calU_{t+1}$;

    \item $S_T \triangleq \SumOneT \zeta_t$ is the cumulative variation of control due to time-varying domain sets. $S_T$ becomes $0$ when domain sets are time-\underline{in}variant; 

    \item $C_T \triangleq \sum_{t=2}^{T} \| u_{t-1}^\star - u_{t}^\star \|$ is the cumulative variation of the optimal control sequence.  It quantifies how fast the optimal control input must  change.
\end{itemize}

\begin{theorem}[Dynamic Policy Regret Bound of \Cref{alg:SafeOGD_Control}]\label{theorem:SafeOGD_Control}
Assume $\eta=\calO\left(\frac{1}{\sqrt{T}}\right)$ for the gradient descend step size.  Then, (i) when \Cref{alg:SafeOGD_Control} is employed to chose online control inputs $u_1, u_2, \ldots$, then it achieves 
\vspace{-2mm}
\begin{equation}
    \DyRegNSC \leq \calO\left(\sqrt{T}\left(1+C_{T}+S_{T}\right)\right),
    \label{eq:theorem_SafeOGD_control_bound}
    \vspace{-2mm}
\end{equation}
against any control inputs $({u}^\star_1, \dots, {u}^\star_T) \in \calU_1 \times \cdots \times \calU_T$.

\noindent (ii) When \Cref{alg:SafeOGD_Control} is employed to choose online control parameters $K_1, K_2, \ldots$, then it achieves 
\vspace{-2mm}
\begin{equation}
    \DyRegNSC \leq \calO\left(\sqrt{T}\left(1+C_{T}+S_{T}\right)\right),
    \label{eq:corollary_SafeOGD_linear_control_bound}
    \vspace{-2mm}
\end{equation}
against any control policies $({K}^\star_1, \dots, {K}^\star_T) \in \calK_1 \times \cdots \times \calK_T$,
where $C_T \triangleq \sum_{t=2}^{T} \| K_{t-1}^\star - K_{t}^\star \|_{\mathrm{F}}$ and $S_T \triangleq \SumOneT \left\|\Pi_{\calK_{t}}(K_{t+1}^\prime) - {K}_{t+1} \right\|_{\mathrm{F}}$.
\end{theorem}

The regret bounds in \Cref{theorem:SafeOGD_Control} reflect the difficulty of the safe non-stochastic control problem.  Ideally, the bounds would be $\calO(\sqrt{T})$, which would imply that \Cref{alg:SafeOGD_Control}
asymptotically guarantees the same performance as the optimal clairvoyant controller because then $\lim_{T\rightarrow\infty} \DyRegNSC/T = 0$.  But the bounds depend on $C_T$ and $S_T$, which in the worst-case {can make the regret linear in $T$:}

\paragraph{Dependency on $S_T$} The bounds in \Cref{theorem:SafeOGD_Control} depend on $S_T$ due to the optimization domain sets being time-varying (\Cref{lemma:u_t}): expectedly, when the safety sets change across time, while the system is threatened by non-stochastic noise, the control becomes more challenging, and, thus, it is more difficult for \Cref{alg:SafeOGD_Control} to match the performance of the optimal controller in hindsight.  In contrast, $S_T$ is zero when the domain sets are time-\underline{in}variant.  
Then, in particular, the regret bounds in \Cref{theorem:SafeOGD_Control} reduce to the near-optimal bounds for the standard \OCO setting with time-\underline{in}variant domain sets:

\vspace{-1.5mm}
    \begin{remark}[Optimality under Time-Invariant Domain of Optimization]\label{remark:bound_reduction_domain}
    When the optimization domain sets are time-\underline{in}variant, that is, $\calU_1 \!=\! \ldots \!=\! \calU_T$ (or $\calK_1 \!=\! \ldots \!=\! \calK_T$), then $S_T = 0$. Therefore, the regret bounds in \Cref{theorem:SafeOGD_Control} reduce to $\calO\left(\sqrt{T}\left(1+C_{T}\right)\right)$, which are  near-optimal~\cite{zhang2018adaptive}, matching the regret bound of the seminal \OGD algorithm for the standard \OCO setting with time-\underline{in}variant constraints~\cite{zinkevich2003online}.\footnote{An example is the case of no safety constraints.}
    \end{remark}
    
\vspace{-2mm}
More broadly, $S_T$ can be sublinear in applications where any two consecutive safe sets differ a little.

\paragraph{Dependency on $C_T$}  The bounds in \Cref{theorem:SafeOGD_Control} depend on $C_T$ due to the optimal control sequence/policy being in general time-varying.
Specifically, \cite{zhang2018adaptive} proved that any optimal dynamic regret bound for \OCO must depend on $C_T$, being lower bounded by $\Omega\left(\sqrt{T(1+C_T)}\right)$.
Expectedly, when the unknown noise sequence requires the system to  adapt its control input frequently, then the harder for \Cref{alg:SafeOGD_Control} is to match the optimal control sequence. 
\vspace{-2mm}
    \begin{remark}[Optimality under also Time-Invariant Control Policies]\label{remark:bound_reduction_control}
        When both the domain sets and the optimal control input sequence in hindsight are time-\underline{in}variant, \ie $S_T=C_T=0$, then the regret bounds in \Cref{theorem:SafeOGD_Control} reduce to $\calO\big(\sqrt{T}\big)$. 
        This implies \SafeOGD converges to the optimal controller in hindsight 
        since then $\lim_{T\rightarrow\infty} \DyRegNSC/T = 0$.
        For example, the classical \scenario{LQG} optimal-control problem has an optimal solution with the form $u_t = -K x_t$, where $K$ is time-\underline{in}variant~\cite{xue2007linear}.
        Thus, if we apply \SafeOGD to learn $K$, then \SafeOGD will converge to an optimal one.
    \end{remark}

%% file: 6-Exp.tex
\vspace{-3mm}
\section{Numerical Evaluations}\label{sec:exp}

\input{Table/Table-Pendulum.tex}

We evaluate \Cref{alg:SafeOGD_Control} in extensive simulated scenarios of safe control, where the controller aims to track a reference setpoint/path while satisfying the state and control input constraints. Particularly, we first consider an inverted pendulum aiming to stay upright despite noise disturbances~(\Cref{subsec:sim-1}) and learn a linear feedback policy.
Then, we consider a quadrotor flying in cluttered environments subject to unknown external forces~(\Cref{subsec:sim-2}) and learn directly the control input. 
Our algorithm is observed in the simulations to (i) achieve comparable or better tracking performance than the linear feedback, \DDPG, geometric control, and \RNMPC, and (ii) guarantee the safety of the system, in contrast to the \iLQR and geometric controllers, which violate the safety guarantees.


\vspace{-3mm}
\subsection{Inverted Pendulum}\label{subsec:sim-1}

\myParagraph{Simulation Setup} We consider an inverted pendulum model with the state vector its angle $\theta$ and angular velocity $\dot{\theta}$, and control input $u$ the torque. The goal of the pendulum is to stay at $(\theta, \dot{\theta}) = (0,0)$. The dynamics of the pendulum are:
\begin{equation}
    \theta_{t+1} = \theta_{t} + \Delta t \dot{\theta}_{t}, \; \dot{\theta}_{t+1} = \frac{3g}{2l} \sin \theta_{t} + \frac{3 \Delta t}{m l^2} u_{t},
\end{equation}
where $g=10~m/s^2$ is the acceleration of gravity, $m=1kg$ is the mass, $l=1m$ is the length of the pendulum, and $\Delta t = 0.05s$ is the time step.
We use: $[-\pi/2 \ -\pi/2]^\top~\leq [\theta_t \ \dot{\theta}_t]^\top~\leq~[\pi/2 \ \pi/2]^\top$ and $-4~\leq~u_{t}~\leq~4$.
We use the loss functions with the form of $c_t(\theta_{t+1}, \dot{\theta}_{t+1}, u_t) = \theta_{t+1}^2 + 0.1 \dot{\theta}_{t+1}^2 + 0.001u_{t}^2$ to learn a linear feedback controller.

We simulate the setting for $T~=~500$ time steps. We corrupt the system dynamics with process noise drawn for the Gaussian, Uniform, or Laplace distribution; and we assume $\| w_{t} \|\leq 0.1$. We perform the simulation $5$ times for each noise distribution in Python with CVXPY solver \cite{diamond2016cvxpy}. 

\myParagraph{Compared Algorithms} 
We compare \Cref{alg:SafeOGD_Control} with: a linear feedback (\LF) controller that stabilizes the linearized pendulum dynamics, the Deep Deterministic Policy Gradient (\DDPG)~\cite{silver2014deterministic,lillicrap2015continuous,stable-baselines3}, and the iterative linear–quadratic regulator (\iLQR)~\cite{chen2017constrained}.
The \LF and \DDPG handle the safety constraints by projecting the control input onto the safe domain set constructed by \Cref{lemma:u_t}; the \iLQR, instead, adds a penalty about the constraint violation to the objective function~\cite{chen2017constrained}.

\myParagraph{Results} The results are given in \Cref{table_pendulum}. 
\Cref{alg:SafeOGD_Control} demonstrates better cumulative loss compared to \DDPG and \LF, achieving 100\% safety rate.
The \iLQR, instead, achieves the lowest loss but only with a $60\%$ safety rate.  

\begin{figure}[t]
	\centering
    \captionsetup{font=footnotesize}
	\includegraphics[width=0.4\textwidth]{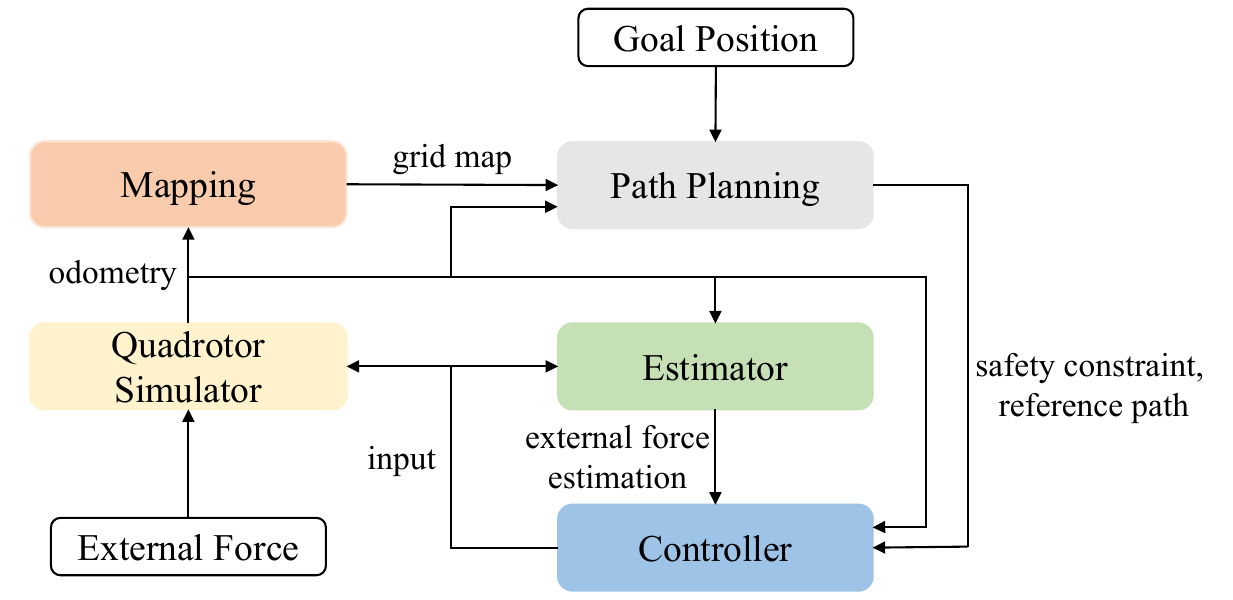}
	\caption{{Autonomous system architecture in \Cref{subsec:sim-2}.}}
    \label{fig_system}
\end{figure} 

\vspace{-5mm}
\subsection{Quadrotor}\label{subsec:sim-2}

\input{Table/Table-Quadrotor}

\begin{figure*}[!]
    \captionsetup{font=footnotesize}
	\centering
	\includegraphics[width=0.95\textwidth]{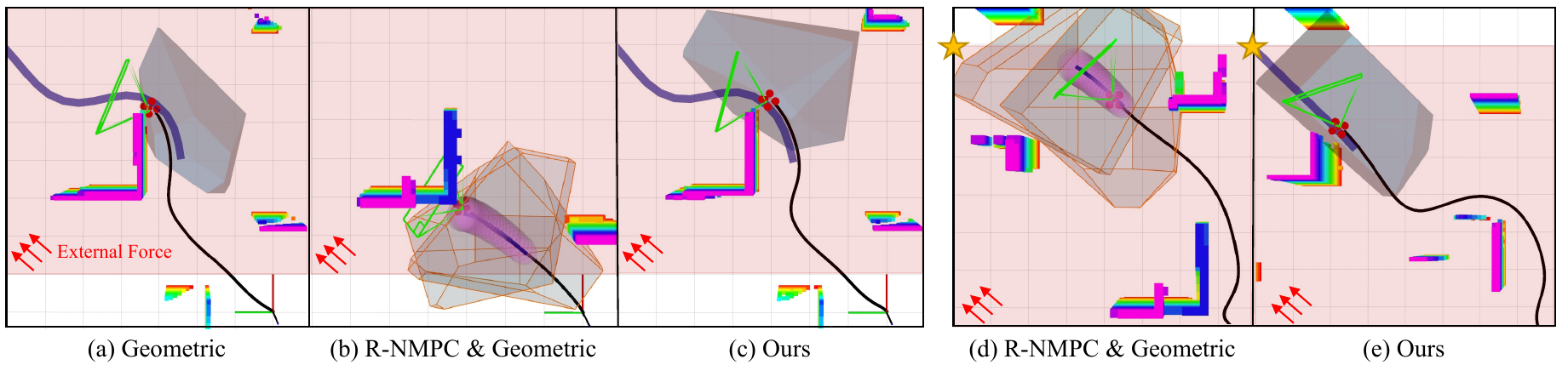}
	\caption{{Simulation results with goal position $\left[10 \ 10 \ 1 \right]^\top$ and disturbances $\left[3 \ 3 \ 0 \right]^\top$ in \Cref{subsec:sim-2}.} The black line is the trajectory, the blue line is the reference trajectory, the red zone is the area where the external forces are applied to the quadrotor, the shaded polytope is the safety constraint, and the gold star is the goal position. (a) \cite{lee2010geometric} collides with obstacles and often has poor safety rate; (b) \cite{wu2021external} collides with obstacles due to latency of \scenariot{\textit{R-NMPC}}; {(d) \scenariot{\textit{R-NMPC}} in \cite{wu2021external} has longer flight time and trajectory length since it aims to guarantee safety against worst-case disturbances over a lookahead horizon; (c) \& (e) Our method achieves collision avoidance while having better performance in flight time, trajectory length, and tracking error.}}
    \vspace{-6.mm}
    \label{fig_quadrotor_rviz}
\end{figure*}

\myParagraph{Simulation Setup} The quadrotor dynamics and the Gazebo simulation environment and system architecture are as follows:
\subsubsection{Quadrotor Dynamics} 
\begin{align}
    \dot{\boldsymbol{p}} &= \boldsymbol{v}, \quad & m \dot{\boldsymbol{v}} &= m \boldsymbol{g} + \boldsymbol{f} + \boldsymbol{f}_w,  \label{eq_uav_trans} \\
    \dot{\boldsymbol{R}} &=\boldsymbol{R}\left[\boldsymbol{\omega}\right]^{\wedge}, \quad & \mathcal{J} \dot{\boldsymbol{\omega}} &= -\boldsymbol{\omega} \times \mathcal{J} \boldsymbol{\omega} + \boldsymbol{\tau}, \label{eq_uav_attitude}
    \vspace{-3mm}
\end{align}
where $\boldsymbol{p} \in \mathbb{R}^{3}$ and $\mathbf{v} \in \mathbb{R}^{3}$ are position and velocity in the iniertial frame, $\boldsymbol{R} \in \mathrm{SO}(3)$ is the attitude rotation matrix, $\boldsymbol{\omega} \in \mathbb{R}^{3}$ is the body angular velocity, $m$ is the quadrotor mass, $\mathcal{J}$ is the inertia matrix of the quadrotor, the wedge operator $^\wedge: \mathbb{R}^3 \rightarrow \mathfrak{s} \mathfrak{e}(3)$ is the skew-symmetric mapping, $\mathbf{g}$ is the gravity vector, $\boldsymbol{f} =  \boldsymbol{R}\left[0\ 0\ T\right]^\top\in \mathbb{R}^3$ and $\boldsymbol{\tau} \in \mathbb{R}^3$ are the total thrust and body torques from four rotors, $T$ is the thrust from four rotors along the $z-$axis of the body frame, and $\boldsymbol{f}_w \in \mathbb{R}^3$ is the unknown external force.

\subsubsection{Gazebo Environment and System Architecture} The Gazebo environment is illustrated in \Cref{fig_gazebo}. We simulate the quadrotor using RotorS~\cite{RotorS2016}. The quadrotor is equipped with an Inertial Measurement Unit (IMU) and a camera.

The system architecture is illustrated in \Cref{fig_system}. We use occupancy grid mapping~\cite{thrun2002probabilistic} for mapping the unknown environment. The external force estimator using VID-Fusion~\cite{ding2021vid} provides an estimate $\tilde{\boldsymbol{f}}_w$ of $\boldsymbol{f}_w$. The path planning module using EGO-Planner~\cite{zhou2021ego} provides a sequence of desired position $\boldsymbol{p}_r$, velocity $\boldsymbol{v}_r$, acceleration $\dot{\boldsymbol{v}}_r$, yaw angle, and yaw rate. We generate polytopic safety constraints using DecompROS~\cite{DecompROS}, {which uses maps of inflated obstacles to account for a quadrotor's size}. We use OSQP solver~\cite{osqp} for the projection step.

\subsubsection{Control Design} Similar to \cite{morgan2015swarm,shi2019neural}, we focus on the translational dynamics in \cref{eq_uav_trans} and use \Cref{alg:SafeOGD_Control} to design the desired force $\boldsymbol{f}_d$, which is then decomposed into the desired rotation matrix $\boldsymbol{R}_d$ and the desired thrust $T_d$ given a desired yaw angle. To this end, we assume that a nonlinear attitude controller, \eg \cite{lee2010geometric}, generates the desired torque such that the desired rotation matrix $\boldsymbol{R}_d$ is tracked. 

The desired force takes the form of $\boldsymbol{f}_d = - k_p \boldsymbol{e}_{\boldsymbol{p}} - k_v \boldsymbol{e}_{\boldsymbol{v}} + m \dot{\boldsymbol{v}}_r - m \boldsymbol{g} - \tilde{\boldsymbol{f}}_w + m \boldsymbol{a}$,
where $\boldsymbol{e}_{\boldsymbol{p}}$ and $\boldsymbol{e}_{\boldsymbol{v}}$ are tracking errors in position and velocity, $k_p$ and $k_v$ are control gains, $\tilde{\boldsymbol{f}}_w$ is the estimation of the external force, and $\boldsymbol{a}$ is the control input learned by \Cref{alg:SafeOGD_Control} using loss function $250 \|\boldsymbol{e}_{\boldsymbol{p}}\|^2 + 10\|\boldsymbol{e}_{\boldsymbol{v}}\|^2$.  
We use the second-order Runge-Kutta (RK2) method~\cite{kloeden1992stochastic} for discretization. 

\subsubsection{Source of Non-Stochastic Noise} Since a stochastic model for the estimation error  $\|\tilde{\boldsymbol{f}}_w - \boldsymbol{f}_w\|$ is generally unknown, we assume the bound $\|\tilde{\boldsymbol{f}}_w - \boldsymbol{f}_w\|\, \leq 0.1$; \ie $f_w=\tilde{f}_w+\mathbf{n}$, where $\|\mathbf{n}\|\,\leq 0.1$, and $\mathbf{n}$ plays the role of non-stochastic noise in the quadrotor dynamics.

\subsubsection{Benchmark Experiment Setup}
We consider that the quadrotor is tasked to fly to prescribed goal positions in the presence of unknown constant external forces that simulate sudden wind gusts; the goal positions and the direction of the wind gusts are specified in \Cref{table_quadrotor}. Particularly, the external forces are applied to the quadrotor as long as its $x$ and $y$ positions are within the box $(x,y) \in [1,\ 10] \times [-10, \ 10]$. We use as performance metrics the flight time, trajectory length, tracking error, computational time, and safety rate.

\myParagraph{Compared Algorithms} 
We compare \Cref{alg:SafeOGD_Control} with \cite{lee2010geometric} and \cite{wu2021external}. \cite{lee2010geometric} is a geometric controller that tracks a desired path based on the feedback of the tracking errors. \cite{wu2021external} uses a robust nonlinear model predictive control (\RNMPC) as a planner with the translational dynamics in \cref{eq_uav_trans} and a simplified attitude dynamics model. Particularly, given the external force estimation from VID-Fusion, the desired acceleration $\dot{\boldsymbol{v}}_r$ provided by \RNMPC in \cite{wu2021external} compensates the external force $\boldsymbol{f}_w$. With the \RNMPC working as a low-frequency outer-loop controller, \cite{wu2021external} then use \cite{lee2010geometric} as a high-frequency inner-loop tracking controller.

\myParagraph{Results} 
 The results are given in \Cref{table_quadrotor} and \Cref{fig_quadrotor_rviz}. 
\Cref{alg:SafeOGD_Control} demonstrates improved performance to \cite{lee2010geometric,wu2021external} in terms of the flight time, trajectory length, tracking error, and safety rate. Moreover, our method is able to achieve $100\%$ safety rate across all tested scenarios, in contrast to \cite{lee2010geometric,wu2021external}. 
The reasons that \cite{lee2010geometric,wu2021external} do not achieve $100\%$ safety rate appear to be the following: \cite{lee2010geometric} has higher tracking error under unknown forces (\Cref{fig_quadrotor_rviz}\!~a) and \RNMPC in \cite{wu2021external} results in a collision due to latency (\Cref{fig_quadrotor_rviz}\!~b). 
Under varying disturbances, the high tracking error of \cite{lee2010geometric} leads to the worst performance in flight time and trajectory length. Additionally, \RNMPC in \cite{wu2021external} has longer flight time and trajectory length than \Cref{alg:SafeOGD_Control} since it aims to guarantee safety against worst-case disturbances over a lookahead horizon (\Cref{fig_quadrotor_rviz}\!~d). 

%% file: Table/Table-Pendulum.tex

\renewcommand{\arraystretch}{1.2} 
\begin{table}[t]
    \captionsetup{font=footnotesize}
    \centering
    \caption{\textbf{Performance Comparison given the Pendulum System in \Cref{subsec:sim-1}.} The table reports the average value and standard deviation of computational time and cumulative loss. The performance is quantified per the computational time, cumulative loss, and safety rate. The \red{red} numbers correspond to the \red{worse} performance.}
     \label{table_pendulum}
     \resizebox{0.85\columnwidth}{!}{
     {
     \begin{tabular}{cccc}
     \toprule
        &  Computational Time ($ms$) & Cumulative Loss & Safety Rate \cr
    \midrule
        Ours  &    $21.6903 \pm 7.8186$   & $20.0022 \pm 1.9933$ & $ 100\% $ \cr
        \DDPG &     $27.6517 \pm 30.7014$  & $26.4953 \pm 1.1261$ & $ 100\% $ \cr 
        \iLQR &    \red{$575.6701 \pm 14.7182$} & $9.3902 \pm 0.5825$  & \red{$ 60\% $}  \cr 
        \LF &    $10.9756 \pm 2.2418$  & \red{$26.6495 \pm 1.2891$} & $ 100\% $ \cr
    \bottomrule
    \vspace{-3mm}
    \end{tabular}}
     }
\end{table}

%% file: Table/Table-Quadrotor.tex
\begin{table*}[t]
    \captionsetup{font=footnotesize}
    \centering
     \caption{\textbf{Performance Comparison for the Quadrotor Experiments in \Cref{subsec:sim-2}.} The table reports the average value and standard deviation of flight time, trajectory length, tracking error, and computational time. The \red{red} numbers correspond to the \red{worse} performance. Our method achieves better performance in terms of flight time, trajectory length, tracking error, and safety rate. \cite{lee2010geometric} has the worst tracking error, resulting in low safety rates. {\cite{wu2021external} either collides due to latency or has longer flight time and length as it aims to guarantee safety against worst-case disturbances over a lookahead horizon}.}
     \label{table_quadrotor}
     \resizebox{1.95\columnwidth}{!}{
     {
     \begin{tabular}{ccccccccc}
     \toprule
       \multirow{2}{*}{Goal Position ($m$)} & \multirow{2}{*}{Disturbance ($m/s^2$)} & \multirow{2}{*}{Method} & \multirow{2}{*}{Flight Time ($s$)} & \multirow{2}{*}{Trajectory Length ($m$)} & \multirow{2}{*}{Tracking Error ($m$)} & \multirow{2}{*}{Safety Rate} &  \multicolumn{2}{c}{Computational Time ($ms$)}  \cr
       \cmidrule(lr){8-9} & & & & & & & Planner & Controller \cr
    \midrule

       \multirow{3}{*}{$\left[\begin{array}{ccc}
           14 & -10 & 1 
       \end{array}\right]^\top$}  & \multirow{3}{*}{$\left[\begin{array}{ccc}
           2  & 2 & 0 
       \end{array}\right]^\top$} &  Ours  &  $9.9720  \pm 0.0944$  &   $18.0600 \pm 0.6542$  &    $0.0811 \pm 0.0130$ & $ 100\% $    & $0.8333 \pm 2.8868$ & \red{$0.1837 \pm 1.3433$}   \cr
       & & \cite{lee2010geometric}  &  $10.1433\pm 0.1914$  &   $18.0333 \pm 0.2309$  &    \red{$0.3022 \pm 0.0078$}  & \red{$ 60\% $} & $0.7692 \pm 2.7735$ & $0.1049 \pm 1.0192$  \cr
       & & \cite{wu2021external}  &  \red{$11.9675 \pm 1.0678 $}  &   \red{$19.0597 \pm 0.7588$}  &    $0.1657 \pm 0.0055$  & $ 80\% $  & \red{$10.0833 \pm 1.8277$} & $0.1091 \pm 1.0392$  \cr
    \midrule
    
       \multirow{3}{*}{$\left[\begin{array}{ccc}
           10 & 10 & 1 
       \end{array}\right]^\top$}  & \multirow{3}{*}{$\left[\begin{array}{ccc}
           3  & 3 & 0 
       \end{array}\right]^\top$} &  Ours  &  $ 8.7340 \pm 0.2050$  &   $ 15.8600 \pm 0.6877$  &    $0.1000 \pm 0.0134$ & $ 100\% $  &  $0.9091 \pm 3.0151$ & \red{$0.1692 \pm 1.2902$}   \cr
       & & \cite{lee2010geometric}  &  $ 10.2625 \pm  0.2869$  &   \red{$ 16.5250 \pm 0.6185 $}   &    \red{$ 0.2694 \pm 0.0056 $}  & \red{$ 40\% $}  & $0.9091 \pm 3.0151$ &  $0.1350 \pm 1.1546$    \cr
       & & \cite{wu2021external}  &  \red{$10.4967 \pm 0.6469$}  &   $16.4467 \pm 0.4520$  &    $0.1874 \pm 0.0391$  & $ 60\% $  & \red{$9.4624 \pm 2.2616$} & $0.1081 \pm 1.0344$  \cr       
    \midrule
    

       \multirow{3}{*}{$\left[\begin{array}{ccc}
           10 & 10 & 1 
       \end{array}\right]^\top$}  & $\left[\begin{array}{ccc} -2  & 2 & 0  \end{array}\right]^\top$ if $x\in [1,4]$, &  Ours  &  $ 8.8788 \pm  0.4825 $  &   $ 15.9125 \pm 0.5111 $  &    $ 0.1156 \pm 0.0080 $ & $ 100\% $  &  $ 0.9091 \pm 3.0151 $ & \red{$ 0.2050 \pm 1.1418$}   \cr
       & $\left[\begin{array}{ccc} -2  & 0 & 0  \end{array}\right]^\top$ if $x\in (4,7)$, & \cite{lee2010geometric}  &   \red{$ 12.1800 \pm 0.5675  $}  &    \red{$ 16.8600 \pm 1.3012$} &    \red{$ 0.2543 \pm 0.0178$}  & $ 100\% $  & $ 0.9091 \pm 3.0151 $ &  $ 0.1243 \pm 1.1085 $    \cr
       & $\left[\begin{array}{ccc} -2  & -2 & 0  \end{array}\right]^\top$ if $x\in [7,10]$ & \cite{wu2021external}  &  $ 10.6400 \pm 0.5915 $  &   $ 16.4388 \pm 0.1008 $  &    $ 0.1889 \pm 0.0244$  & \red{$ 60\% $}  & \red{$ 9.6109 \pm 2.3058 $} & $ 0.1053 \pm 1.0212 $  \cr       

    \bottomrule
    \end{tabular}}
    }
    \vspace{-4mm}
\end{table*}

%% file: 7-Conclusion.tex
\vspace{-4mm}
\section{Conclusion} \label{sec:con}

\myParagraph{Summary} We studied the problem of \textit{Safe Non-Stochastic Control for Control-Affine Systems} (\Cref{prob:SafeNSC}), and provided the \SafeOGD algorithm. \SafeOGD  guarantees safety and bounded dynamic regret against any time-varying control policy (\Cref{theorem:SafeOGD_Control}). 
\SafeOGD is near-optimal for classical online convex optimization and optimal control problems:  
    (i) when the domain of optimization is time-\underline{in}variant,
    \SafeOGD's performance bound reduces to the bound of the \OGD algorithm for online convex optimization, which is near-optimal~\cite{zinkevich2003online};
    and (ii) when also the optimal clairvoyant control policy is time--\underline{in}variant, \SafeOGD can learn it asymptotically, implying, for example, that \SafeOGD will converge to the optimal linear state-feedback controller if applied to the classical \scenario{LQG} problem.
We evaluated our algorithm in simulated scenarios of an inverted pendulum aiming to stay inverted, and a quadrotor flying in an unknown cluttered environment (\Cref{sec:exp}). We observed that our method demonstrated better performance, guaranteeing safety, whereas state-of-the-art methods, such as the \iLQR~\cite{chen2017constrained} and \RNMPC~\cite{wu2021external} did not.

\myParagraph{Future work} We will investigate the regret bound of the \SafeOGD algorithm under unknown \textit{stochastic noise}, providing \textit{Best-of-Both-Worlds} guarantees in mixed stochastic and non-stochastic environments.

%% file: Acknowledgements.tex
\vspace{-1mm}
\section*{Acknowledgements}
We thank Yuwei Wu from the GRASP Lab, University of Pennsylvania, for her invaluable discussion on the numerical simulations of the quadrotor experiment.

%% file: Appendix/Appendix.tex
\vspace{-1mm}
\appendix
\input{Appendix/Appendix-Safe-NSC.tex}

%% file: Appendix/Appendix-Safe-NSC.tex
\vspace{-2mm}
\subsection{Proof of \Cref{theorem:SafeOGD_Control}}\label{app:theorem_SafeOGD_control}
We present the proof for the basic approach. The proof for learning a linear controller follows the same steps.

We define $\bar{u}_{t+1} \triangleq \Pi_{\calU_{t}}(u_{t+1}^\prime)$ and $\zeta_t \triangleq \left\|\bar{u}_{t+1} - {u}_{t+1} \right\|$. By convexity of $c_t$, we have
\begin{equation}
    \begin{aligned}
        & c_t\left(u_t\right)-c_t\left(u^\star_t\right) \\
        \leq& \left\langle\nabla c_t\left(u_t\right), u_t-u^\star_t\right\rangle=\frac{1}{\eta}\left\langle  u_t-u_{t+1}^{\prime}, u_t-u^\star_t\right\rangle \\
        =& \frac{1}{2 \eta}\left(\left\|u_t-u^\star_t\right\|^2-\left\|u_{t+1}^{\prime}-u^\star_t\right\|^2+\left\|u_t-u_{t+1}^{\prime}\right\|^2\right) \\
        \leq& \frac{1}{2 \eta}\left(\left\|u_t-u^\star_t\right\|^2-\left\|\bar{u}_{t+1}-u^\star_t\right\|^2\right)+\frac{\eta}{2} G^2 ,
        \label{eq:theorem_SafeOGD_1}
    \end{aligned}
\end{equation}
where the last inequality holds due to the Pythagorean theorem \cite{hazan2016introduction}, \Cref{assumption:bounded_set}, and \Cref{assumption:cost}.

Consider the term $\left\|\bar{u}_{t+1}-u^\star_t\right\|^2 = \left\|{u}_{t+1}-u^\star_t\right\|^2 + \left\|u_{t+1}-\bar{u}_{t+1}\right\|^2 - 2 \left\langle {u}_{t+1}-u^\star_t,  u_{t+1}-\bar{u}_{t+1} \right\rangle$, we have
\begin{equation}
    \begin{aligned}
        & c_t\left(u_t\right)-c_t\left(u^\star_t\right) \\
        \leq& \frac{1}{2 \eta} \Bigl( \left\|u_t-u^\star_t\right\|^2 - \left\|{u}_{t+1}-u^\star_t\right\|^2 - \left\|u_{t+1}-\bar{u}_{t+1}\right\|^2 \\
        & \qquad + 2 \left\|{u}_{t+1}-u^\star_t\right\| \left\|{u}_{t+1}-\bar{u}_{t+1}\right\| \Bigr) +\frac{\eta}{2} G^2 \\
        \leq& \frac{1}{2 \eta}\left(\left\|u_t-u^\star_t\right\|^2 - \left\|{u}_{t+1}-u^\star_t\right\|^2 \right) + \frac{D \zeta_t}{\eta} +\frac{\eta}{2} G^2 \\
        =& \frac{1}{2 \eta}\left(\left\|u_t\right\|^2 - \left\|{u}_{t+1}\right\|^2 \right) +\frac{1}{\eta} \left(u_{t+1}-u_t\right)^{\top} u^\star_t + \frac{D \zeta_t}{\eta} +\frac{\eta}{2} G^2,
        \label{eq:theorem_SafeOGD_3}
    \end{aligned}
\end{equation}
where the second inequality holds due to $\left\|u_{t+1}-\bar{u}_{t+1}\right\|^2 \geq 0$, $\left\|{u}_{t+1}-u^\star_t\right\| \leq D$ by \Cref{assumption:bounded_set}, and $\zeta_t \triangleq \left\|\bar{u}_{t+1} - {u}_{t+1} \right\|$.

Summing \cref{eq:theorem_SafeOGD_3} over all iterations, we have 
\begin{equation}
    \begin{aligned}
        & \sum_{t=1}^T c_t\left(u_t\right) - \sum_{t=1}^T c_t\left(u^\star_t\right) \\
        \leq& \frac{1}{2 \eta}\left\|u_1\right\|^2 + \frac{1}{\eta}\left(u_{T+1}^{\top} u^\star_T-u_1^{\top} u^\star_1\right) \\
        &\qquad +\frac{1}{\eta} \sum_{t=2}^T\left(u^\star_{t-1}-u^\star_t\right)^{\top} u_t + \frac{D}{\eta} \sum_{t=1}^T\zeta_t + \frac{\eta T}{2} G^2 \\
        \leq& \frac{7 D^2}{4 \eta} + \frac{D}{\eta} C_T + \frac{D}{\eta} S_T + \frac{\eta T}{2} G^2,
        \label{eq:theorem_SafeOGD_4}
    \end{aligned}
\end{equation}
where the last step holds due to \Cref{assumption:bounded_set} and the Cauchy-Schwarz inequality, \ie $\left\|u_1\right\|^2 \leq D^2$, $u_{T+1}^{\top} u^\star_T \leq\left\|u_{T+1}\right\|\left\|u^\star_T\right\| \leq D^2$, $-u_1^{\top} u^\star_1 \leq \frac{1}{4}\left\|u_1-u^\star_1\right\|^2 \leq \frac{1}{4} D^2$, $\left(u^\star_{t-1}-u^\star_t\right)^{\top} u_t \leq\left\|u^\star_{t-1}-u^\star_t\right\|\left\|u_t\right\| \leq D\left\|u^\star_{t-1}-u^\star_t\right\|$, along with the definition of $C_T$ and $S_T$.

Choosing $\eta=\calO\left(\frac{1}{\sqrt{T}}\right)$ gives the result in \cref{eq:theorem_SafeOGD_control_bound}.
\qed



\vspace{-3mm}
\subsection{Proof of \Cref{lemma:cvx_u}}\label{app:lemma_cvx_u}
The proof follows by the convexity of $c_{t}\left(x_{t+1},u_t\right): \mathbb{R}^{d_{x}} \times \mathbb{R}^{d_{u}} \mapsto \mathbb{R}$ in $x_{t+1}$ and $u_t$ by \Cref{assumption:cost}, and the linearity of $x_{t+1}$ in $u_t$, \ie $x_{t+1} =  f\left(x_{t}\right) + g\left(x_{t}\right) u_{t} + w_{t}$, given functions $f\left(\cdot\right)$ and $g\left(\cdot\right)$, $x_t$, and $w_t$.
\qed

\vspace{-3mm}
\subsection{Proof of \Cref{lemma:u_t}}\label{app:lemma_u_t}
Consider at time step $t$, we aim to choose $u_t$ such that 
\begin{align}
        x_{t+1} &= f\left(x_t\right) + g\left(x_t\right) u_t + w_t  \nonumber \\
        &\in \calS_{t+1} \triangleq \{x \mid L_{x,t+1} x \leq l_{x,t+1}\}, \ \forall {w}_{t} \in \calW, \label{eq:lemma_u_t_0} \\
        u_t &\in \calC_t \triangleq \{u \mid L_{u,t} u \leq l_{u,t}\},  \label{eq:lemma_u_t_1}
\end{align}
given $f\left(\cdot\right)$, $g\left(\cdot\right)$, $x_t$, $L_{x,t+1}$, $l_{x,t+1}$, $L_{u,t}$, and $l_{u,t}$.

The second constraint on control input, $u_t \in \calC_t$, can be directly imposed. We now consider the constraint on state $x_{t+1}$ and define the control barrier function at time step $t$ and $t+1$: $h_{t}\left(x\right) = l_{x,t} - L_{x,t} x, \ h_{t+1}\left(x\right) = l_{x,t+1} - L_{x,t+1} x$.


The DCBF condition in \cref{eq:DCBF} gives
\vspace{-1.5mm}
\begin{equation}
    \begin{aligned}
        \Delta h_t + \alpha h_{t}\left(x_t\right) &=  - L_{x,t+1} g\left(x_{t}\right) u_{t} - L_{x,t+1} w_{t} + l_{x,t+1} \\
                    & \quad  - L_{x,t+1} f\left(x_{t}\right)  - (1-\alpha)\left(l_{x,t} - L_{x,t} x_t\right) \\
                    &\geq 0, \ \forall {w}_{t} \in \calW .
    \end{aligned}
    \label{eq:lemma_u_t_4}
\vspace{-1.5mm}
\end{equation}

By applying robust optimization \cite{ben2009robust}, \cref{eq:lemma_u_t_4} becomes
\vspace{-1.5mm}
\begin{equation}
    \begin{aligned}
        \Delta h_t + \alpha & h_{t}\left(x_t\right) \geq  - L_{x,t+1} g\left(x_{t}\right) u_{t} - \left\| L_{x,t+1} \right\| W + l_{x,t+1} \\
                    & \quad  - L_{x,t+1} f\left(x_{t}\right)  - (1-\alpha)\left(l_{x,t} - L_{x,t} x_t\right) \geq 0 .\\
    \end{aligned}
    \label{eq:lemma_u_t_5}
\vspace{-1.5mm}
\end{equation}

By combining \cref{eq:lemma_u_t_1,eq:lemma_u_t_5}, we construct $\calU_t$ as
\begin{equation}
    \begin{aligned}
    & \calU_t  \triangleq \{  u \mid \  - L_{x,t+1} g\left(x_{t}\right) u - \left\| L_{x,t+1} \right\| W + l_{x,t+1} \\
                & \ \  L_{u,t} u \leq l_{u,t}, \ - L_{x,t+1} f\left(x_{t}\right) - (1-\alpha)\left(l_{x,t} - L_{x,t} x_t\right) \geq 0 \},
    \end{aligned}
    \label{eq:lemma_u_t_11}
\end{equation}
which is convex in $u_t$. Choosing $u_t \in \calU_t$ ensures that the safety constraints in \cref{eq:lemma_u_t_0,eq:lemma_u_t_1} are satisfied.
\qed